# Breast cancer detection using artificial intelligence techniques: A systematic literature review

Ali Bou Nassif*, Manar Abu Talib, Qassim Nasir, Yaman Afadar, Omar Elgendy
{anassif, mtalib, nasir, u17104387, u16104886}@sharjah.ac.ae
University of Sharjah, UAE

* Corresponding Author

## Abstract
Cancer is one of the most dangerous diseases to humans, and yet no permanent cure has been developed for it. Breast cancer is one of the most common cancer types. According to the National Breast Cancer foundation, in 2020 alone, more than 276,000 new cases of invasive breast cancer and more than 48,000 non-invasive cases were diagnosed in the US. To put these figures in perspective, 64% of these cases are diagnosed early in the disease's cycle, giving patients a 99% chance of survival. Artificial intelligence and machine learning have been used effectively in detection and treatment of several dangerous diseases, helping in early diagnosis and treatment, and thus increasing the patient's chance of survival. Deep learning has been designed to analyze the most important features affecting detection and treatment of serious diseases. For example, breast cancer can be detected using genes or histopathological imaging. Analysis at the genetic level is very expensive, so histopathological imaging is the most common approach used to detect breast cancer. In this research work, we systematically reviewed previous work done on detection and treatment of breast cancer using genetic sequencing or histopathological imaging with the help of deep learning and machine learning. We also provide recommendations to researchers who will work in this field.

## 1. Introduction

Breast cancer is one of the major causes of death in women around the world. According to the American cancer society, 41,760 women and more than 500 men died from breast cancer recently[1]. Breast cancer occurs in four main types: normal, benign, in-situ carcinoma and invasive carcinoma [1]. A benign tumor involves a minor change in the breast structure. It is not harmful and does not classify as a harmful cancer. In cases of in-situ carcinoma, the cancer is only in the mammary duct lobule system and does not affect other organs. This type is not dangerous and can be treated if diagnosed early. Invasive carcinoma is considered to be the most dangerous type of breast cancer, as it can spread to all other organs. According to the authors in [1], breast cancer can be detected using several methods including X-ray mammography, ultrasound (US), Computed Tomography (CT), Portion Emission Tomography (PET), Magnetic Resonance Imaging (MRI) and breast temperature measurement. Usually, the golden standard is a pathological diagnosis for detecting breast cancer. This involves an image analysis of the removed tissue, which is stained in the lab to increase visibility. Hematoxylin and Eosin (H&E) are commonly used for the staining process. Breast cancer can be diagnosed using one of two approaches: histopathological image analysis or genomics. Histopathological images are microscopic images of breast tissue that are extremely useful in early treatment of the cancer. As for genomics, the authors in [2] stated that radio-genomics is an emerging research field focusing on multi-scale associations between medical imaging and gene expression data.

---

[1] https://www.cancer.org/

Radio-genomics provide both radiological and genetic features that can enhance diagnosis. It can analyze tissues at the molecular level, helping with prediction and early detection of cancer. The main difference between imaging information and radio-genomics is the critical knowledge gap between imaging at the tissue level and analyzing the underlying molecular and genetic disease biomarkers. As imaging is less precise, it may lead to over- or under-treatment. While radio-genomics is much more effective than histopathological imaging, it is rarely used because the process involves datasets that are very expensive and require high computational power. As a result, a limited number of labs conduct radio-genomics experiments [2]. This research paper addresses the following research questions and highlights the deep learning models, looking at their performance, the datasets used and possibilities for breast cancer classification and detection.

1. **Which deep learning models perform most effectively?**
   We will compare deep learning models with classical machine learning models to compare their performance. We will also list the performance metrics used.

2. **What are the most used features for breast cancer classification? How are these features selected and extracted?**
   We will observe the most important features that contribute to breast cancer classification, and the methods used to extract these features.

3. **What datasets are available for both gene sequencing and MRI? What feature selection and extraction methods are used?**

   We will list and discuss all public and private datasets for gene sequencing and MRI imaging data. We will also list some of the methods used to select and extract the features.

4. **Comparing gene sequence data with image data, for breast cancer detection problem, what are the drawbacks, challenges, and advantages?**
   We will compare imaging and gene sequencing as they relate to breast cancer detection, using a tabular presentation to highlight the main differences between the two approaches.

The remaining of this paper is structured as follows: Section 2 presents related work which includes surveys conducted in breast cancer area. Section 3 explains the methodology used to conduct this research. Section 4 presents the obtained results and related discussions. Lastly, Section 5 concludes the paper and suggests future research directions.

## 2. Related Work

Many studies have been conducted about breast cancer detection through imaging or through genomics. However, to the best of our knowledge, no research has been conducted including both techniques.

The authors in [1] summarized the various techniques used to classify breast cancer using histopathological image analysis (HIA) based on different architectures of Artificial Neural Networks (ANN). The authors grouped their work according to the applied dataset. They arranged it in ascending chronological order. This work found that ANNs were first used in the field of HIA around 2012. ANNs and PNNs were the most frequently applied algorithms. However, in feature extraction, most of the work used textural and morphological features. It was clear that Deep CNNs were quite effective for early detection and diagnosis of breast cancer, leading to more successful treatment. Prediction of Non-Communicable Diseases (NCDs) was conducted using many algorithms.

In [2], the authors compared the performance of various classification algorithms. The classification algorithms were performed on eight NCD datasets using eight classification algorithms and a 10-fold cross-validation method. These were evaluated using AUC as an indicator of accuracy. The authors stated that the NCD datasets have noisy data and irrelevant attributes. KNN, SVM and NN proved to be robust to this noise. In addition, they stated that the irrelevant attribute problem can be handled with some pre-processing techniques to improve the accuracy rate.

Natural inspired computing (NIC) algorithms have been designed and applied to diagnose various human disorders. The authors in [3] introduced five insect-based NIC algorithms used for diagnosing diabetes and cancer. The authors found that it achieved a high level of performance in detecting different types of cancer (breast, lung, prostate and ovarian). To be more specific, breast cancer was detected using a hybridization of the guided ABC and neural networks.

The authors also developed a highly effective methodology of detecting diabetes and leukemia. They concluded that the hybridization of NICs with other classification algorithms produces more precise and promising results. They mentioned that more work is required to detect different stages of diabetes and cancer.

In [4], the authors demonstrated the effectiveness of NNs in the classification of cancer diagnoses, especially in the initial stages. According to their study, the majority of NNs have shown promise in detecting tumor cells. However, the imaging approach requires high computational capacity to preprocess the images.

In [5], the authors reviewed different machine learning, deep learning and data mining algorithms related to breast cancer prediction. Several research papers on breast cancer were reviewed, with a total of 27 papers in machine learning, 4 papers in ensemble techniques and 8 papers in deep learning techniques. The authors mentioned that most of the papers used imaging, while only a few papers used genetics. The main algorithms used to predict breast cancer using genetics were SVM, decision tree and random forest. However, imaging techniques used several algorithms such as CNNs and Naïve Bayes.

On the other hand, the authors in [6] focused on gene mutation for detecting breast cancer. They mentioned that the gene prediction classification phase aims to carry out gene annotation, gene finding and gene mutation detection to ascertain the presence or absence of a cancer. They concluded that several methods can be used including regression, probability models, SVMs, NNs and deep learning. They also mentioned the many opportunities available to capture the relationship between nucleotide and feature extraction, since DNA sequencing involves a large amount of data in the form of a string sequence.

In [7], the authors examined recent studies applying deep learning to breast cancer with different imaging modalities. They organized these studies using the aspects of dataset, architecture, application and evaluation. They focused on deep learning frameworks developed in three breast imaging modalities (ultrasound, mammography and MRI). In their work, they attempted to provide state-of-the-art findings about breast cancer imaging utilizing DLR-based CAD systems. Their study included private datasets and classification using CNNs.

After studying these surveys, our contribution will involve studying genetic sequencing and imaging at the same time to predict breast cancer and to get more information that can help early diagnosis and treatment of breast cancer. We will also provide recommendations to researchers who wish to conduct research in this area.

## 3. Methodology

Our target topic is breast cancer detection using deep learning. We ended up using around 80 of the most recent papers related to breast cancer treatment and diagnosis. Some of the papers examined only deep learning, while others used a combination of machine learning and deep learning.

In our search process, we mainly used the Scopus database to obtain the articles. This is to exclude non-refereed publications However, in Figure 1, we state the distribution of selected papers among the existing databases. The top five databases are PubMed, ScienceDirect (Elsevier), IEEE, Springer and Nature.

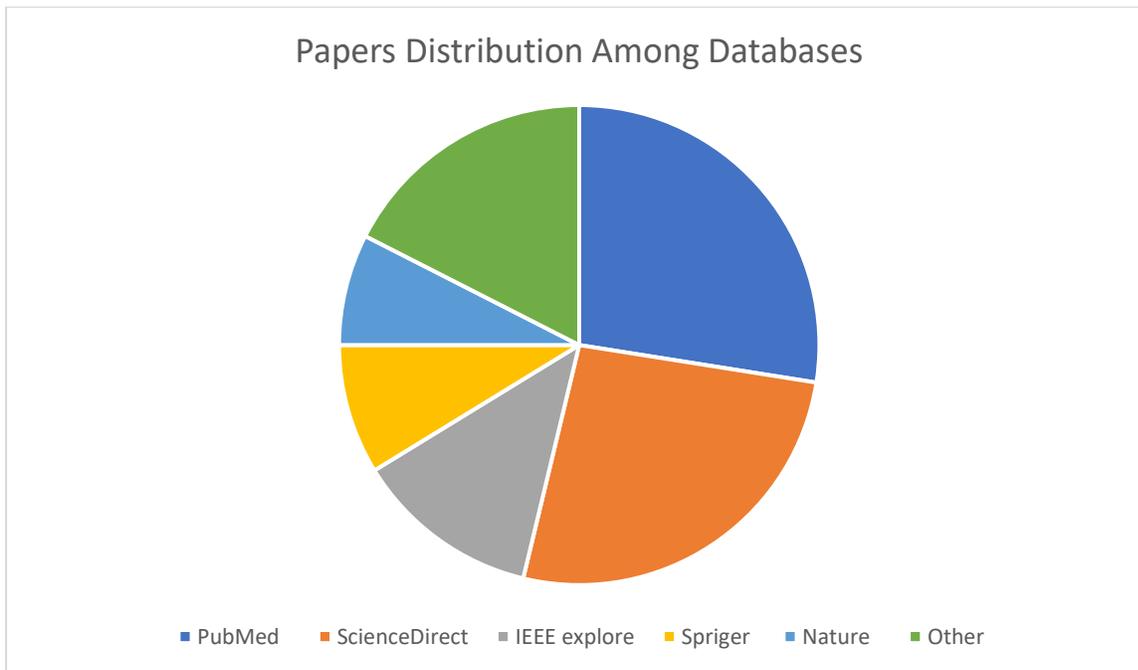

*Figure 1 Papers distribution among databases*

We used the following search statement: ("breast cancer") AND ALL ("deep learning" OR "deep neural network") AND ALL ("gene" OR "genome" OR "microarray" OR "DNA" OR "X-ray" OR "mammography" OR "MRI" OR "ultrasound"). More than 1,000 papers were found that were published between January 2010 and May 2020. The following figure shows the distribution of paper publication during this period.

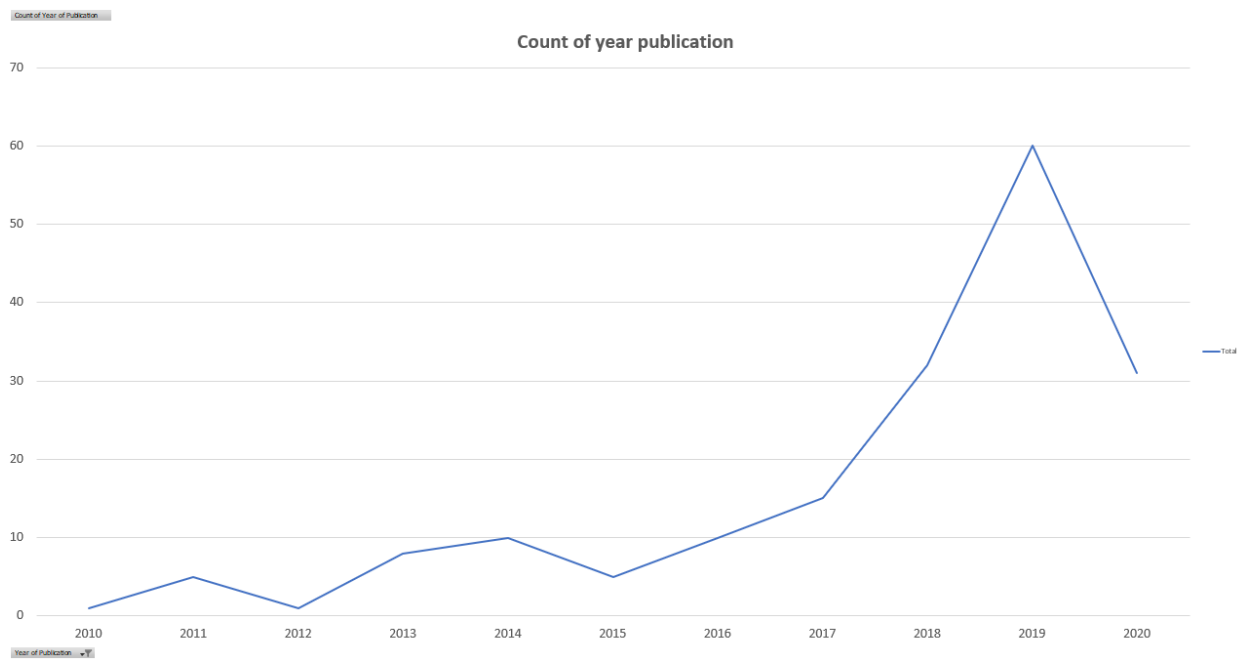

*Figure 2 Number of publications between 2010 and 2020.*

We can see that breast cancer study publications peaked around 2019 and 2020. We reduced the number of papers to 80 by including only papers using genetic expression and imaging and by focusing on journal and conference articles only. The imaging modalities we considered were ultrasound, radiography, mammography and magnetic resonance imaging (MRI), as well as various types of gene expression and gene sequencing. In this research, we focused on papers that implement the breast cancer detection using the techniques of AI, as well as papers that predict breast cancer using both gene data and image data. We applied the following eligibility criteria on each paper: (1) The language is English; (2) The topic is related to breast cancer detection and treatment; (3) The paper discusses hybrid models of machine learning and deep learning; (4) The paper purely discusses deep learning; (5) The paper discusses genetic expression data; (6) The paper discusses imaging data; (7) Only journal and conference publications are retained; (8) Only medical or biomedical engineering publications are kept which are related to the topic. Please note that non-refereed publications were excluded from the study.

Firstly, we recorded the main information such as the name of the paper, year of publication, the list of authors and the publisher. Then, we included some information to conduct the systematic review, such as the algorithm used and whether the paper discusses only deep learning or a hybrid between DL and ML, the recorded accuracy and other performance evaluation parameters, the dataset, the features, and many other columns. We answered our research questions using these criteria.

## 4. Results and Discussion

In this systematic study, our initial search turned up 1,000 conference and journal papers. After eliminating duplicated papers and unrelated studies that were "purely medical or about cancer in general", we ended up with 80 papers related to both ML and DL.

We wanted to focus on DL approaches or DL-ML hybrid models, so only papers related to DL were selected. Figure 3 explains our search methodology.

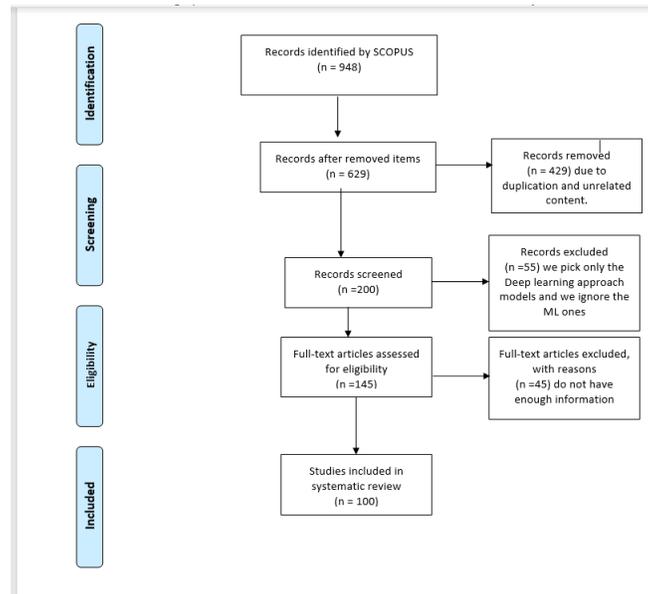

*Figure 3 Flow of information through the phases of a systematic review.*

1. **Which deep learning models perform most effectively?**

The following tables (Tables 1 & 2) summarize the algorithms used and their performance, with some details related to both genetic and imaging data.

*Table 1 Models, classes, and performance for gene sequencing data in selected papers.*

| Paper Reference | Models/ Algorithm | Binary or Multiclass | Classes | Accuracy | Other Performance Evaluation Parameters | Anomaly Application/ Task |
|---|---|---|---|---|---|---|
| [8] | IABC-EMBOT, IHM-FFNN, PSO-RM, ABCO-BCD and DNN-BCD | Binary | Negative, positive | 0.975 | _ | BC detection |
| [9] | FNN, ANFIS, ANNFIS | Binary | Negative, positive | 0.92 | Precision: 0.944, recall: 0.944, F1:0.944 | BC detection |
| [10] | Deep type, state-of-the-art | Multiclass | Normal, luminal A, luminal B, basal and HER2 | _ | _ | Identifying cancer subtypes |
| [11] | BPNN | Binary | Mutant and non-mutant sequences | 0.998 | Sensitivity=1 Specificity=0.9985 | Predication and Classification of Cancer |
| [12] | CNN | Multiclass | - | 0.956 | - | BC subtype classification |
| [13] | DA | Multiclass | - | 0.95 | - | BC detection |
| [14] | DNN + Attention mechanism (Hybrid) | Binary | - | 0.87 | - | BC detection |
| [15] | GCN | Multiclass | - | 0.919 | AUC=0.84 | Synergistic drug combinations |
| [16] | DNN+SVM (separately) | Multiclass | Binary, Miotic/non | 0.94 | F-score: 0.556 Accuracy: 0.8319 | Detection |
| [17] | DNN | Multiclass | 4 classes - Basal-like, HER2-enriched, Luminal A, and Luminal B And binary (Basal, Non basal) | 0.83 | AUC: 0.82 Accuracy: 0.8682 | Identify Risk categories |
| [18] | DL + ML (Separately) CNN, SVM, Random Forests, Boosting | Multiclass | Axillary lymph node status, binary, cancer or not | 0.97 by SVM | Accuracy: 0.98 AUC: 0.93 | Cancer subtype classification |
| [19] | DL + ML (Hybrid) Genetic Algorithm (GA) based MLP, Multilayer Perceptron (MLP), Logistic Regression (LR) | Multiclass | Binary, cancer or not | 0.84 | AUC: 0.84 | Prediction of axillary lymph node status in breast cancer |
| [20] | Feed Forward Neural Network (FFNN) | Binary | Binary, cancer or not | 0.983 | Only accuracy | BC detection |
| [21] | DNN | Multiclass | Cancer types | 0.64 | - | BC detection |
| [22] | CNN | Binary | Tumor or not | 0.967 | Only accuracy | |
| [23] | CNN | Multiclass | 7 Cancer types | 0.846 | Only accuracy | BC subtypes identification |
| [24] | PCA, PCA-AE-Ada | Binary | Binary | 0.85 | SN=0.84m SP=0.55, AUC=0.74 | Predict clinical outcome of breast cancer |
| [25] | Deeptriage | Multiclass | Normal, luminal A, luminal B, basal and HER2 | - | F1- score= 0.90 | BC subtypes identification |
| [26] | CNN+HASHI | Multiclass | - | - | NPV=0.97, TNR=0.92, FPR=0.08, FNR=0.13, PPV=0.72, Dice=0.76 | BC detection |
| [27] | CNN, GSS | Binary | Binary | - | Recall = 0.66 Precision = 0.826 | Mitosis detection |

| | | | | | F-score = 0.734 | |
|---|---|---|---|---|---|---|
| [28] | CNN | Both | Binary and (epithelium, stroma, and fat) | 0.92 and 0.95 | - | Identify and classify tumor-associated stroma in diagnostic breast biopsies. |
| [29] | SVM, NB, RF | Binary | ER- and ER+ | 0.85 | - | Cancer sub-types (ER+ and ER−). |
| [30] | RF, ANN | Multiclass | 7 classes | - | - | Gene prediction |
| [31] | Deep CNN | Binary | Epithelial and stromal | 0.88 | - | Epithelial and stromal |
| [32] | DNN | Multiclass | Basal-like, HER2-enriched, Luminal A, and Luminal B | 0.87 | - | BC Molecular subtype classification |
| [33] | ANN, LR cascaded and individually | Binary | Death/ Survival | 0.84 | Sensitivity, specificity, AUC, various for different number of years and features | BC survival according to different features |
| [34] | DNN | Multiclass | - | 0.82 | Precision = 0.875 Sensitivity = 0.200 Matthew's correlation coefficient = 0.356 | BC prognosis detection |
| [35] | AP + ANN | Binary | Malignant/ benign | 0.983 | Sensitivity = 0.9803 Specificity = 0.9887 | Feature selection method |
| [36] | CNN | Multiclass | - | - | AUC = 0.902 Dice coefficient = 0.7586 | BC detection |
| [37] | PNN | Binary | Malignant/ benign | 0.963 | Sensitivity = 0.9888 | BC early diagnosis |
| [38] | ANN | Binary | Malignant/ benign | 0.989 | - | BC diagnosis |
| [39] | CNN, SVM | Multiclass | 4 tissue categories | 0.9 | F-score = 0.94 AUC = 0.99 | BC detection |
| [40] | CNN, Autoencoder | Binary | Positive/ Negative | 0.986 | Sensitivity = 0.9812 Specificity = 0.9877 Precision = 0.9688 | BC diagnosis |

| | | | | | F score = 0.9750 | |
|---|---|---|---|---|---|---|
| [41] | KNN, ANN, SVM, LS-SVM | Binary | Positive/ Negative | 0.953 | - | BC image classification |
| [42] | ANN | Regression | Tumor weight | - | R^2 = 0.897 RMSE= 0.271 | BC tumor detection |
| [43] | SVM, CNN, SLIC | Binary | Epithelial and non-epithelial | 0.942 | Precision = 0.9283 Recall = 0.992 F score = 0.958 | BC image classification |
| [44] | ANN | Multiclass | ER/HER2-, ER/HER2+, ER-/HER2+, ER-/HER2- | - | - | BC tumor detection |
| [45] | DNN | Nuclei probability | - | - | F score = 0.59±0.14 Precision = 0.72±0.12 Recall = 0.56±0.2 Specificity = 0.9±0.06 | BC tumor detection |
| [46] | ANN, KNN, RBFNN, SVM | Binary | Malignant/ benign | 0.965 | F score = 0.962 Specificity = 0.938 Sensitivity = 0.960 | BC detection |
| [47] | Pre-trained networks: VGG16, VGG19, and ResNet50 | Binary | Malignant/ benign | 0.926 | ROC = 0.956 Precision = 0.959 | BC detection |
| [48] | NNC | Binary | Healthy/Cancer | 0.971 | AUC = 0.991 Sensitivity = 0.957 Specificity = 0.976 | BC detection |
| [49] | CNN, Hierarchical Classification | Binary | Malignant/ benign | 0.954 | Sensitivity = 0.935 | BC detection |
| [50] | CNN, SVM | Binary | Low risk, high risk | - | (95% CI 1.33–3.32, p=0.001) | BC risk outcome |
| [51] | CNN | Binary and Multiclass | Malignant/ benign, subclasses | 0.98 | F score = 0.843 Precision = 0.842 | BC tumor detection |
| [52] | ANN, NB | Binary | Malignant/ benign | 0.98 | - | BC tumor detection |
| [53] | ANN + (Correntropy + Hinge + Cross-Entropy) | Binary | Malignant/ benign | 0.97 | Precision = 0.1 Recall = 0.94 | BC tumor detection |

| | | | | | F score = 0.97 | |
|---|---|---|---|---|---|---|
| [54] | ANN, Clustering | Binary | Malignant/ benign | 0.96 | - | BC tumor detection |
| [55] | DBN, SVM | Multiclass | Normal, UC, UD, CD | 0.90 | - | BC gene classification |
| [56] | Constructive DNN | Multiclass | Low risk/ intermediate risk/high risk | 0.87 | NPV = 0.92 PPV = 0.48 TNR = 0.93 TPR = 0.50 | BC risk outcome |
| [57] | LR, DT, ANN, SVM, KNN, GNB, RF | Multiclass | Luminal A/ luminal B/ HER2-Enriched/ basal-like | 0.95 | - | BC subtype identification |
| [58] | DNN, Clustering | Multiclass | Luminal A/ luminal B/ HER2-Enriched/ basal-like | - | - | BC subtype identification |
| [59] | CNN | - | - | - | F score = 0.652 Spearman = 0.617 Kappa = 0.567 | BC detection |
| [60] | BAT, GSA, FNN | Binary | Malignant/ benign | 0.942 | Recall = 0.943 Specificity = 0.893 Precision = 0.8403 | BC tumor detection |
| [61] | CNN, SVM | - | - | 0.840 | AUC = 0.852 Sensitivity = 0.867 Specificity = 0.833 | BC lymph node detection |
| [62] | NN | Multiclass | ER/PR/Her2 | 0.950 | AUC = 0.890 | BC tumor detection |
| [63] | CNN | Binary | Normal/tumor | 0.987 | Sensitivity = 0.914 Specificity = 0.100 Precision = 0.100 F score = 0.955 | BC tumor classification |
| [64] | FCM, PCA, GLCM, KNN, SVM | Binary | Malignant/ benign | - | - | BC tumor detection |
| [65] | Meta + SVM, Metalogistic regression, Integrative deep learning | Multiclass | - | - | Sensitivity = 0.8284 Specificity = 0.9815 | BC gene identification |
| [66] | SVM, RF | Binary | Malignant/ benign | 0.970 | F score = 0.950 | BC tumor detection |
| [67] | single-output Chebyshev- | Binary | Malignant/ benign | 0.994 | - | BC tumor detection |
| | polynomial neural network (SOCPNN), and the modified SOCPNN | | | | | |
| [68] | RF, CT, LR, MLP, LSTM, GRU | Multiclass | - | - | F score = 0.820 | Tumor subclass detection |
| [69] | CNN | Multiclass | - | 0.945 | Precision = 0.958 Recall = 0.956 F score = 0.964 | BC detection |
| [70] | MLP, RNN | Binary | Malignant/ benign | - | AUC = 0.998 F score = 0.980 | BC detection |
| [71] | CNN | Binary | Normal/cancer | 0.956 | Time = 80.3 seconds | BC type detection |
| [72] | DNN | Multiclass | ER-/ER+/Triple negative | - | Recall = 0.950 Precision = 0.900 F score = 0.920 | BC tumor detection |

In the previous table, generally, the work in this area is divided into two main groups, the first involving binary classification (whether or not breast cancer is present) and the second classifying breast cancer types. We noticed that the use of binary classification results in the highest accuracy, and that it is generally more accurate than multiclass classification. In paper [11], the author used the BPNN algorithm in a highly effective way. We also see that most selected papers with binary classifications performed well in terms of most evaluation parameters. [18] However, the best performance for multiclass categorization or breast cancer subtypes classification obtained 97% accuracy. The author of the paper basically compared machine learning and deep learning on the task of breast cancer subtype classification. Most of the papers were strictly focused on accuracy and did not mention other parameters such as precision, recall, AUC, etc. This is an important limitation when we are talking about medical projects. Many models have been used, including CNN, DNN with attention mechanism, Feed Forward Neural Network (FFNN) and many other DL mechanisms. Because not all papers took into consideration the confusion matrix parameters, we will include only accuracy in our graphs.

To state the findings from the above table, CNN model seems to be the number one used model among the papers for both binary and multiclass classification. A hybrid between machine learning and deep learning also seems to be affective for such models. We can see a hybrid between MLP and LR. As well as SVM algorithm from ML has been used a lot in these papers.

Basically, when it comes to breast cancer detection using imaging data, we see high performance on binary classification. However, when it comes to classification of types, imaging did not produce results as accurately as gene expression data. Paper [77] obtained very high accuracy using a hybrid model between DL and ML for breast cancer detection.

For multiclass differentiation or breast cancer subtype classification, the highest accuracy obtained using imaging data was 90% in a paper that uses ML. We can therefore see in the above table that in general, the use of CNN results in excellent performance for both gene expression and imaging. An example of this is paper [75], which produced 97% accuracy for binary classification. The models used include both deep learning standalone models and hybrid models consisting of both machine learning and deep learning algorithms.

Comparing between hybrid and deep learning standalone models in gene expression, we can see that the standalone deep learning models obtain consistently higher accuracy. The hybrid technique in paper [14] obtained 87% accuracy, which was the lowest accuracy in all the models. The CNN in paper [12] obtained an accuracy of 95%, while the algorithm BPNN achieved 99.8% accuracy. Many papers did not mention confusion matrix parameters. In fact, very few papers mentioned these parameters in which the authors mentioned sensitivity and specificity. Among the imaging papers, the highest accuracy was 99.7% for binary classification in paper [80], which used a hybrid model with both ML and DL.

*Table 2 Model, classes, and best performance for MRI imaging data*

| Paper Reference | Models/ algorithm Image | Binary or Multiclass | Classes | Accuracy | Other performance evaluation parameters | Anomaly application |
|---|---|---|---|---|---|---|
| [73] | CNN | Multiclass | Luminal A Luminal B HER2 | 0.70 | ROC=0.85 | BC subtype classification |
| [74] | CNN | Multiclass | Complete, partial, no response | 0.88 | Specificity of 0.951, sensitivity of 0.739 | Predict breast tumor, Response to chemotherapy |
| [75] | CNN | Binary | Negative and positive | 0.972 | Sensitivity 0.983, and Specificity 0.965 | BC detection |
| [76] | ANN, NB, K-NN, DT, RF | Multiclass | - | 0.90 | - | BC subtype classification |
| [77] | CNN | Binary | Negative and positive | 0.873 | | |
| [78] | DL + ML (Hybrid) Logistic regression, random forest and deep neural network | Binary | Negative and positive | 0.98 | - | BC detection |
| [79] | 2CNN, 3CNN | Binary | Negative and positive | 0.705 | AUC=0.763, sensitivity=0.805, specificity=0.618 | BC detection |
| [80] | DL + ML (Hybrid) Model 1: DBN-ELM- BP Model 2: DBN-BP-ELM Model 3: DBN + GA | Binary | Negative and positive | 0.9975 | - | BC detection |

According to the above table, the mostly used algorithm for breast cancer detecting and subtype classification is CNN, and that does make sense, since our data is MRI images and CNN considered the best for computer vision problems.

To conclude, many algorithms were used in the studies. Some papers used several models in series, while others used only one model. According to Figure 4, ANN and CNN were the most widely used algorithms in both gene sequence and images data. Many other algorithms were used, such as DNN and SVM, but most of the papers used CNN and ANN with various parameters and properties.

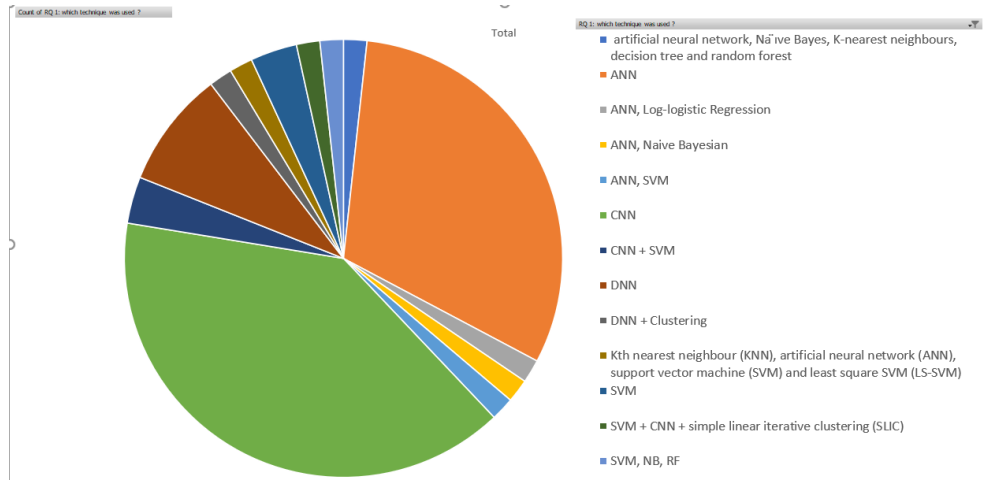

*Figure 4 Most commonly used algorithms in the papers*

According to Figures 5 and 6, models with the highest accuracies are SOCPNN and CNN, respectively.

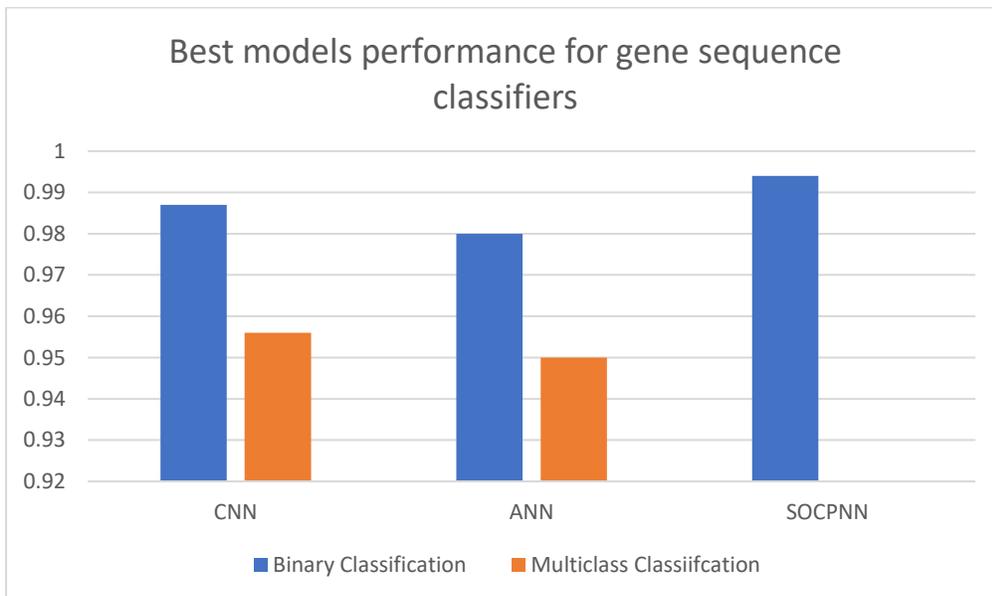

*Figure 5 Best models performance for gene sequence classifiers*

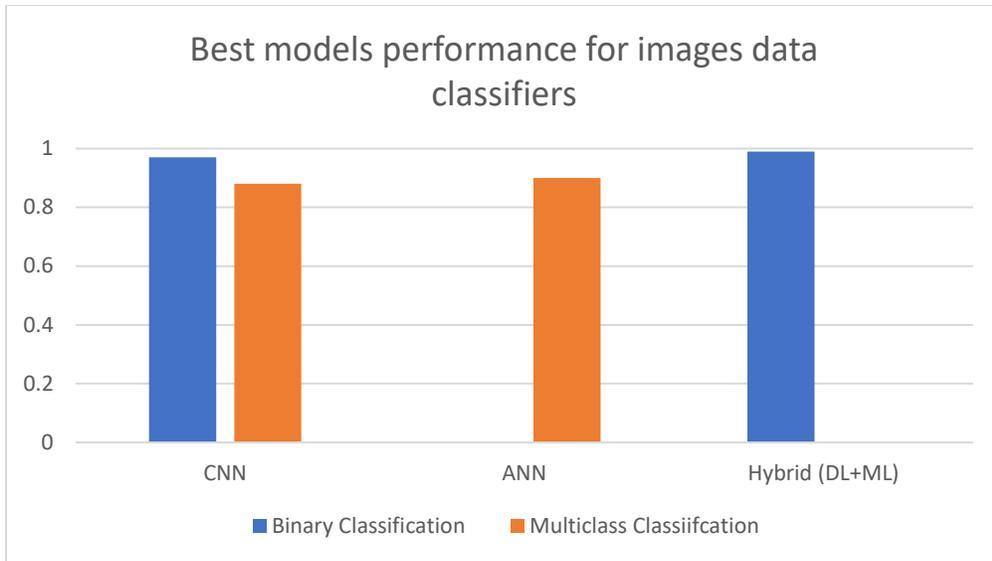

*Figure 6 Best models performance for images data classifiers*

### 2. What datasets are available for gene sequencing and MRI?

According to the figure below, many public and private datasets were used. The private datasets were sourced from various universities in the US and EU.

There were very few publicly available datasets that were not for free. Table 3 shows the public and private datasets available for gene sequencing.

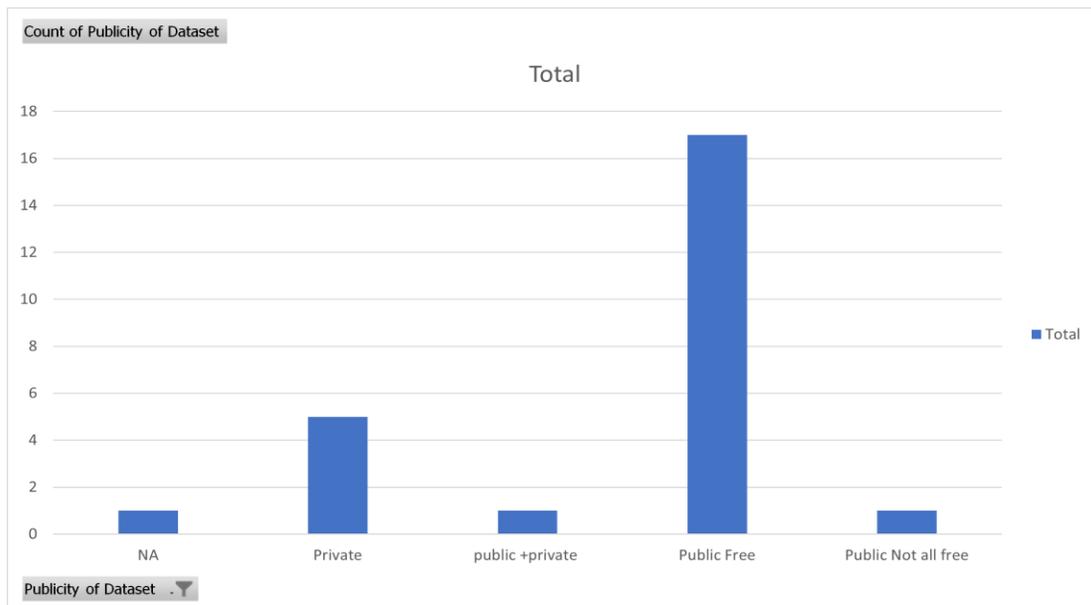

*Figure 7 Dataset publicity*

*Table 3 Public and private datasets available for gene sequencing*

| Paper reference | Dataset | Publicity | availability | Link | # of instances |
|---|---|---|---|---|---|
| [16] [13] [81] | The cancer Genome Atlas | Public | Free | Http://cancergenome.nih.gov | 11429 |
| [18] | METABRIC datasets | Public | Free | Https://ega-archive.org/datasets/EGAD00010000268 Https://www.cbioportal.org/study/summary?Id=brca_metabric | 543 |
| [82] | Array express database | Private | - | - | - |
| [83] | GEO database | Public | | Https://www.ncbi.nlm.nih.gov/geo/info/download.html | 404 |
| [15] | STRING and BIOGRID | Private | Free | - | - |
| [84] | NCI Genomic Data Commons (GDC) | Public | Paid | Https://gdc.cancer.gov/ | 9114 |
| [85] | Spark dataset | Public | Free | Https://drive.google.com/file/d/1yd1gwk2owgoooq9wi1k7puoakd7cbs8t/view | 106 |

Although gene expression data is not as common as imaging data, we found many public and private datasets containing gene expression data for both healthy and sick people. The most well-known of these is the cancer Genome Atlas [86], which is a project aiming to identify the complete set of DNA changes in many different types of cancer. Studying these changes may help researchers understand how different types of cancer form. The dataset contains gene data for many cancer types, including breast cancer. The second most commonly used dataset is the METABRIC dataset, which contains clinical traits, expression, CNV profiles and SNP genotypes derived from breast tumors collected from participants of the METABRIC trial. The GEO database contains genome data and DNA sequencing from cancer detection research. NCI Genomic Data Commons (GDC) also provides researchers with a large number of gene related data for cancer research and analysis.

The most used dataset in the above table is the cancer Genome Atlas dataset. It provides researchers with a big number of instances. It provides a clinical data for each participants including some general information.

Table 4 explains the available imaging datasets for detecting breast cancer.

*Table 4 Public and private datasets for MRI imaging*

| Paper reference | Dataset | Publicity | Cost | Link | No of instances |
|---|---|---|---|---|---|
| [87], [80] | Wisconsin breast cancer dataset | Public | Free | Https://archive.ics.uci.edu/ml/datasets/Breast+Cancer+Wisconsin+(Diagnostic) | 569 |
| [31] | Helsinki University & Netherlands Cancer Institute & Vancouver General Hospital | Private | - | - | - |
| [88] | MRI dataset | Public | Free | Https://wiki.cancerimagingarchive.net/display/Public/RIDER+Breast+MRI#2251275749b786f1af5747c39abd8eda0d12e2b7 | 1500 |
| [89] | University of Vermont Medical Center | Private | - | - | |
| [90] | Digital Database for Screening Mammography (DDSM) | Public | Free | Https://wiki.cancerimagingarchive.net/display/Public/CBIS-DDSM#225166295e40bd1f79d64f04b40cac57ceca9272 | 10239 |
| [91] | Stanford Tissue Microarray Database (TMA) | Private | - | - | - |
| [92] | Mammographic Image Analysis Society (MIAS) | Public | Free | Http://peipa.essex.ac.uk/benchmark/databases/index.html | 322 |

More datasets are available for imaging data than for genetic data. The most used imaging dataset is the Wisconsin breast cancer dataset, which is obtained from the UCI repository. It contains features that are computed from a digitized image of a Fine Needle Aspirate (FNA) of a breast mass.

Many other breast cancer imaging datasets could be found, most of them public and free. If the dataset is very large, like the DDSM dataset, it can be used on its own. If the dataset is not large enough to produce high performance results, it can be merged with another dataset to improve the data pool.

Researchers mainly used Wisconsin breast cancer dataset and MRI dataset, as we can notice both of them are public and contain a lot of samples.

### 3A. What are the most commonly used features for breast cancer classification?

We can see that the features are divided into two categories: tumor information and protein types/status. The tumor information includes the size of the tumor and the grade, which depends on cancer type. The ER and PR status represent the proteins that exist in the area. Number of positive lymph nodes represents how many lymph nodes contain the cancer and is recorded as a real number. Metastasis records the number of sites the cancer has spread into, and may include the lungs, bones, etc. HER2 is another protein that indicates the growth of the cancer inside the breast area. PAM50 is an index or test that shows if breast cancers are likely to metastasize (spread to other organs). The following tables (Tables 5 & 6) show the features, their description and types for gene as well as imaging.

*Table 5 Genetic data features*

| Feature | Description | Feature type |
| --- | --- | --- |
| ER status | Positive or negative | Nominal |
| PR status | Positive or negative | Nominal |
| Tumor size | In inches | Numeric |
| Tumor grade | Grade A, B and C | Nominal |
| Number of positive lymph nodes | Real number (counts) | Nominal |
| Metastasis | Number of metastatic sites. Count variable | Numeric |
| Age | Real number | Numeric |
| The Nottingham Prognostic Index | Calculated using formula given set parameters. | Numeric |
| PAM50 index | Prognosis test shows the breast cancer has a fairly high risk of metastasis (the PAM50 score is high) | Numeric |
| HER2 | This protein promotes the growth of cancer cells | Nominal |

*Table 6 Imaging data features*

| Feature | Description | Feature type |
| --- | --- | --- |
| Size of epithelial cells | In inches | Numeric |
| Density of epithelial cells | Real number | Numeric |
| Clump thickness | Real number | Numeric |
| Symmetry | 0 or 1 | Nominal |
| Uniformity of cell size | 0 or 1 | Nominal |
| Marginal adhesion | 0 or 1 | Nominal |
| Concave points | 0 or 1 | Nominal |
| Compactness | 0 or 1 | Nominal |

When looking at imaging data, many general features are considered such as symmetry, compactness and concave points. Another type of feature is related to the breast image specifically. For example, marginal adhesion quantifies the degree to which cells on the outside of the epithelial wall tend to stick together. Another example is uniformity of cell size, which is represented by 0 if lacking uniformity and 1 if uniform.

Clump thickness describes if cells are mono- or multi-layered. Uniformity of cell size evaluates consistency in cell size in the sample.

### 3B. What are the most effective feature selection and feature extraction methods?

## Feature extraction methods

The literature discusses various methods for feature extraction. Feature-based Strategy (FES) is a method used in some of the papers for finding image displacements. This strategy pinpoints features (for example image edges, corners and other structures that can be localized in two dimensions) and tracks these as they move from frame to frame. This involves two stages. Firstly, the features are found in two or more consecutive images. The act of feature extraction, if done correctly, will reduce the amount of information to be processed (and so reduce the workload). It will also contribute to obtaining a higher level of understanding in the field, by its very nature of eliminating the unimportant information. Secondly, these features are matched between the frames. In the simplest and commonest cases, two frames are used and two sets of features are matched to produce a single set of motion vectors. Alternatively, the features in one frame can be used as seed points from which to use other methods (i.e.: gradient-based methods -- see the following section) to find the motion [23].

Other authors used a multi-level wavelet transformation method, which is a wavelet-based discrete signal analysis method that can extract multilevel time-frequency features from a time series by decomposing the series level by level into low and high frequency sub-series.

CNN was also used for feature selection and extraction. This is a very broad question, but you can look at general CNN architecture for two main types of imaging classification: "feature extractor" based on convolutional layers, and "classifier", usually based on fully connected layers.

Feature extraction usually refers to one of two options: either the last hidden layer – i.e.: the last layer before the output layer [see vgg16 example below, the 4096x1x1 layer], or the last convolutional layer after flattening [in vgg16 -the conv5 layer 7x7x512] [93].

## Feature selection methods

Several feature selection methods were used, starting with XGboost. The authors in [94] used XGboost and a random forest multilayer network analysis of mRNA and protein expression profiles in breast cancer patients. Random forests are often used for feature selection in a data science workflow because the tree-based strategies used by random forests naturally ranks features by how well they improve the purity of the node. This means that they assign each feature a status on whether they reduce impurity over all trees (called GINI impurity). Nodes producing the greatest decreases in impurity occur at the start of the trees, while nodes with the smallest reduction in impurity occur at the end of trees. Thus, by pruning trees below a particular node, we can create a subset of the most important features.

On the other hand, authors in [83] used principal component analysis (PCA), a technique for reducing the dimensionality of datasets, increasing interpretability but at the same time minimizing information loss. They achieved this by creating new uncorrelated variables that successively maximize variance.

### 4. Comparing gene sequence data with image data, for breast cancer detection problem, what is the drawbacks, challenges, and advantages for each?

In this survey, we compare between genetic data and imaging data as they are related to breast cancer detection. Here, we summarize the findings and results.

*Table 7 The differences between genetic sequencing data and imaging data*

|  | **Image** | **Gene** |
|---|---|---|
| **Features** | More features, most obtained using CNN | Fewer features but more effective |
| **Performance** | Best accuracy is 0.993 | Best accuracy is 0.998 |
| **Advantage** | Easy to use CNN, more available datasets | More accurate, more confidence |
| **Disadvantage** | Many related and nonrelated features. | Hard to obtain enough datasets, possibly expensive, complex. |
| **Medical confidence** | Most common | A focus of recent research. |

We collected papers that use either genetic sequencing data or imaging data to detect breast cancer or to classify it into subtypes. Both strategies produced good results. Regarding imaging data, many features could be easily extracted, but not all of them were effective. Genetic expression data contains fewer features but may be more effective. The best performance using imaging data in our systematic review produced 99.3% accuracy, whereas the most accurate use of genetic data resulted in 99.8% accuracy.

In general, dealing with imaging data is much easier than dealing with genetic data. There are more ways to preprocess imaging data, and more techniques to extract features from it. On the other hand, in our survey, we found that genetic data is consistently more accurate, especially in the context of multiclass prediction.

Imaging data requires more data cleaning work after a feature's extraction phase because most techniques involve CNNs extracting related and nonrelated features, possibly leading to poor performance.

On the other hand, dealing with genetic data poses more challenges. Genetic data processing is complex and expensive. It is much easier to find imaging data for breast cancer projects. When it comes to genetic data, it is hard to find datasets that are large enough with the right labels.

We cannot conclude that gene expression will give better prediction and more accurate results, based on the above comparison we can state that each datatype contains drawbacks and challenges.

## 5. Conclusions and Future Research Directions

Most papers published in the field of breast cancer detection and subtype classification use machine learning techniques. However, deep learning models have not been heavily investigated in this domain. This presents researchers with opportunities to use various deep learning mechanisms to predict patient status such as LSTM, GAN and RNN, as these types of research have not yet been conducted in the field.

Moreover, most papers focus only on the Accuracy metric to evaluate their performance and ignore confusion matrix parameters and AUC. This is insufficient since the Accuracy metric does not distinguish between false positive and false negative classifications. Future studies should include at least AUC and F-scores to assess the performance of each model.

An important finding has been noticed which is the wide use of the CNN algorithm for both gene expression and MRI images data. Such models often obtain good results in comparison to other algorithms. Researchers might be interested to do further searching and implement more hybrid algorithms with CNN.

Furthermore, we noticed that the Attention mechanism has not been used very often to classify images. So, this gives future researchers an opportunity to use Attention to improve the accuracy of deep learning models.

Recently researchers are focusing on the gene sequence data as it is a wide area and there is always room for further research and results.

There are several opportunities for future researchers to contribute by merging multiple gene sequencing datasets to predict additional outcomes with larger dataset.

Future researchers may also focus on extracting significant features from genetic expression data to obtain better results, using confusion matrix parameters to increase accuracy. Additionally, researchers can focus on feature selection methods to eliminate non-significant features, leading to better performance.

Most of the research we found, focused on breast cancer detection and subtype classification. This leaves room for future research to address various related topics such as identifying risk levels and predicting the possibility of recurrence. One direction for future research is related to implementing multiclass predictors using genetic data. Most research papers used genetic sequencing data only with binary classification, with the main focuses being breast cancer detection and likelihood of survival.

**Compliance with Ethical Standards**

The authors thank the University of Sharjah for supporting this work.

Funding is through the competitive project "Applications of Machine Learning in Metastatic Breast Cancer Detection"

Conflict of Interest: The authors declare that they have no conflict of interest.

Informed Consent: This study does not involve any experiments on animals.

## Appendix 1

| Acronym | Explanation |
|---------|-------------|
| ANN | Artificial Neural Network |
| AI | Artificial intelligence |
| BC | Breast cancer |
| HIA | Histopathological image analysis |
| CNNs | Convolution Neural Network |
| NCDs | Non-Communicable Diseases |
| KNN | K Nearest neighbor |
| AUC | Area Under the curve |
| SVM | Support vector machine |
| NIC | Natural inspired computing |
| US | Ultrasound |
| CT | Computed Tomography |
| PET | Portion Emission Tomography |
| MRI | Magnetic Resonance Imaging |
| H&E | Hematoxylin and Eosin |
| DNA | Deoxyribonucleic Acid |
| DL | Deep learning |
| FNN | Feed forward network |
| BPNN | Back propagation neural network |
| GA | Genetic Algorithm |

| PCA | Principle component analysis |
|---|---|
| NB | Naïve bayes |
| LR | Linear regression |
| DT | Decision trees |
| RF | Random Forest |
| RNN | Recurrent Neural network |
| LSTM | Long short-term memory |
| MLP | Multilayer perception |
| ROC | receiver operating characteristic |
| NN | Neural Network |